\documentclass[journal]{IEEEtran}
\usepackage{latexsym,amsmath,amsfonts}
\ifCLASSINFOpdf
\else
\fi
\begin{document}
%
\title{A Combinatorial Necessary and Sufficient Condition for Cluster Consensus}
%
%
%

\author{Yilun~Shang
\thanks{Y. Shang is with the Department
of Mathematics, Tongji University, Shanghai 200092, China (e-mail: shyl@tongji.edu.cn).}

\thanks{Manuscript received August 24, 2015; revised xxxx xx, xxxx.}}

\maketitle

\begin{abstract}
In this technical note, cluster consensus of discrete-time linear
multi-agent systems is investigated. A set of stochastic matrices
$\mathcal{P}$ is said to be a cluster consensus set if the system
achieves cluster consensus for any initial state and any sequence of
matrices taken from $\mathcal{P}$. By introducing a cluster
ergodicity coefficient, we present an equivalence relation between a
range of characterization of cluster consensus set under some mild
conditions including the widely adopted inter-cluster common
influence. We obtain a combinatorial necessary and sufficient
condition for a compact set $\mathcal{P}$ to be a cluster consensus
set. This combinatorial condition is an extension of the avoiding
set condition for global consensus, and can be easily checked by an
elementary routine. As a byproduct, our result unveils that the
cluster-spanning trees condition is not only sufficient but
necessary in some sense for cluster consensus problems.
\end{abstract}


%
\IEEEpeerreviewmaketitle

%
%
%
%

\section{Introduction}

In the past two decades, consensus problems in multi-agent systems
have gained increasing attention in various research communities,
ranging from formation of unmanned air vehicles to data fusion of
sensor networks, from swarming of animals to synchronization of
distributed oscillators \cite{1,2}. The main objective of consensus
problems is to design appropriate protocols and algorithms such that
the states of a group of agents converge to a consistent value (see
\cite{4,5} for a survey of this prolific field). In many distributed
consensus algorithms, the agents update their values as linear
combinations of the values of agents with which they can
communicate:
\begin{equation}
x_i(t+1)=\sum_jp_{ij}(t+1)x_j(t),\label{1}
\end{equation}
where $x_i(t)$ is the value of agent $i$ and $P(t)=(p_{ij}(t))$ for
every discrete time instant $t\ge0$ represents a \textit{stochastic}
matrix, i.e., $p_{ij}(t)\ge0$ and $\sum_jp_{ij}(t)=1$. The states of
agents following such linear averaging algorithms tend to get closer
over time. The problem of characterizing the complete sequence of
matrices $P(t)$ for consensus is however known to be notoriously
difficult \cite{7}. A moderate goal would be to determine whether
the system (\ref{1}) converges to a state of consensus for all
sequences of matrices $P(t)$ in a certain set $\mathcal{P}$.
Remarkably, Blondel and Olshevsky in a recent work \cite{6}
presented an explicit combinatorial condition, which is both
necessary and sufficient for consensus of (\ref{1}) in that sense.
This condition, dubbed ``avoiding sets condition'', is easy to check
with an algorithm and thus the consensus problem is decidable.

While most existing works are concerned with global consensus
(namely, all the agents reach a common state), in varied real-world
applications, there may be multiple consistent states as agents in a
network often split into several groups to carry out different
cooperative tasks. Typical situations include obstacle avoidance of
animal herds, team hunting of predators, social learning under
different environments, coordinated military operations, and task
allocation over the network between groups. A possible solution is
given by the cluster (or group) consensus algorithms \cite{8,9},
where the agents in a network are divided into multiple subnetworks
and different subnetworks can reach different consistent states
asymptotically. Evidently, cluster consensus is an extension of
(global) consensus. Various sufficient conditions and necessary
conditions (although much fewer) for cluster consensus have been
reported in the literature for discrete-time systems \cite{9,10,11},
simple first- or second-order continuous-time systems
\cite{8,12,13,14,15}, and high-order dynamics \cite{16}, to name a
few. However, most of these conditions rely on either complicated
linear matrix inequalities or algebraic conditions involving
eigenvalues of the system matrices, which are in general difficult
to check. Elaborate algebraic necessary and sufficient conditions
have also been obtained for multiconsensus problems, which appear to
be a wider class of cluster consensus \cite{17}.

With the above inspiration, we aim to work on efficiently verifiable
conditions for cluster consensus by extending the results in
\cite{6} for global consensus, which are highly non-trivial. The
main contribution of this paper is to establish a combinatorial
necessary and sufficient condition which guarantees the cluster
consensus of system (\ref{1}) under some common assumptions (such as
the self-loops and the inter-cluster common influence). Some of the
previous convergence criteria can be quickly reproduced from our
results. It is noteworthy that the authors in \cite{9} showed that,
under some mild assumptions, the cluster consensus of (\ref{1}) can
be achieved provided the graphs associated with $P(t)$ contain
cluster-spanning trees. Interestingly, our result implies that the
cluster-spanning trees condition is essentially necessary.

\section{Preliminaries}

In this section, some definitions and lemmas on graph theory and
matrix theory are given as the preliminaries. We refer the reader to
the textbooks \cite{5,18} for more details.

Let $G=(V,E)$ be a directed graph of order $n$ with the set of
vertices $V=\{1,\cdots,n\}$ and the set of edges $E\subseteq V\times
V$. For a stochastic matrix $P=(p_{ij})\in\mathbb{R}^{n\times n}$
(namely, $p_{ij}\ge0$ and $\sum_{j=1}^np_{ij}=1$ for all $i,j$), a
corresponding directed graph $G(P)=(V,E)$ can be constructed by
taking $V=\{1,\cdots, n\}$, and $E=\{(i,j):p_{ij}>0\}$. Given a
subset $S\subseteq V$, denote by $N_G(S)$ the set of out-neighbors
of $S$ in $G$, i.e., $N_G(S)=\{j\in V: i\in S, (i,j)\in E\}$. A
directed path from vertex $i$ to $j$ of length $l$ is a sequence of
edges $(i,i_1), (i_1,i_2), \cdots, (i_l,j)$ with distinct vertices
$i_1,\cdots, i_l\in V$. If $(i,i)\in E$, then there exists a
self-loop at vertex $i$.

A \textit{clustering}
$\mathcal{C}=\{\mathcal{C}_1,\cdots,\mathcal{C}_K\}$ of the directed
graph $G$ is defined by dividing its vertex set into disjoint
clusters $\{\mathcal{C}_k\}_{k=1}^K$. In other words, $\mathcal{C}$
satisfies $\cup_{k=1}^K\mathcal{C}_k=V$ and
$\mathcal{C}_k\cap\mathcal{C}_{k'}=\emptyset$ for $k\not=k'$.
Letting $x(t)=(x_1(t),\cdots,x_n(t))^T$, we recast the system
(\ref{1}) as
\begin{equation}
x(t+1)=P(t+1)x(t).\label{2}
\end{equation}

\noindent\textbf{Definition 1.} For a given clustering
$\mathcal{C}=\{\mathcal{C}_1,\cdots,\mathcal{C}_K\}$, a set of
$n\times n$ stochastic matrices $\mathcal{P}$ is said to be a
\textit{cluster consensus set} if for any initial state $x(0)$ and
all sequences $P(1),P(2),\cdots\in\mathcal{P}$,
\begin{equation}
\lim_{t\rightarrow\infty}x(t)=\sum_{k=1}^K\alpha_k{\bf1}_{\mathcal{C}_k},\label{3}
\end{equation}
where ${\bf1}_{\mathcal{C}_k}$ is the sum of $i$th $n$-dimensional
basis vector $e_i=(0,\cdots,0,\overset{i
\mathrm{th}}{1},0,\cdots,0)^T$ over all $i\in\mathcal{C}_k$, and
$\alpha_{k}$ is some scalar.

\noindent\textbf{Remark 1.} In most of the literatures, given a
sequence of stochastic matrices (or switching signal)
$P(1),P(2),\cdots$, the system (\ref{2}) is said to achieve
\textit{cluster consensus} if (\ref{3}) holds for any initial state
$x(0)$. This is also referred to as \textit{intra-cluster
synchronization} in \cite{9,14,19}, where the cluster consensus
requires additionally the separation of states of agents in
different clusters. Nevertheless, the inter-cluster separation can
only be realized by incorporating adapted external inputs.

\noindent\textbf{Definition 2.} \cite{9} A stochastic matrix $P$ is
said to have \textit{inter-cluster common influence} if for all
$k\not=k'$, $\sum_{j\in\mathcal{C}_{k'}}p_{ij}$ is identical with
respect to all $i\in\mathcal{C}_k$.

\noindent\textbf{Remark 2.} Since the entries on each row of $P$ sum
up to one, the above statement automatically holds for $k=k'$ if $P$
has inter-cluster common influence. Therefore,
$\sum_{j\in\mathcal{C}_{k'}}p_{ij}$ depends only on $k$ and $k'$.
This (and some closely related variants) is a common assumption for
cluster consensus problems; see e.g. \cite{8,9,10,11,12,19}. It is
direct to check that if $P_1$ and $P_2$ have inter-cluster common
influence with respect to the same clustering $\mathcal{C}$, so does
$P_1P_2$.

To analyze the cluster consensus of the multi-agent system
(\ref{2}), we will need to estimate some characteristics of infinite
product of stochastic matrices. For a stochastic matrix
$P=(p_{ij})\in\mathbb{R}^{n\times n}$, we define the cluster
ergodicity coefficient with respect to a clustering $\mathcal{C}$ as
\begin{align}
\tau_{\mathcal{C}}(P)=&\ \frac12\max_{1\le k\le
K}\max_{i,j\in\mathcal{C}_k}\sum_{s=1}^n|p_{is}-p_{js}|\nonumber\\
=&\ \frac12\max_{1\le k\le
K}\max_{i,j\in\mathcal{C}_k}\|p_i-p_j\|_1,\label{4}
\end{align}
where $p_i=(p_{i1},\cdots,p_{in})$ is the $i$th row of $P$ and
$\|\cdot\|_1$ represents the 1-norm of vector.

It can be seen that $0\le\tau_{\mathcal{C}}(P)\le1$ and that
$\tau_{\mathcal{C}}(P)=0$ if and only if
$P=\sum_{k=1}^K{\bf1}_{\mathcal{C}_k}y_k^T$, where $y_k$ is a
stochastic vector, namely, $P$ has identical rows for each cluster.
Hence, $\tau_{\mathcal{C}}$ can be viewed as an extension of the
well-known Dobrushin ergodicity coefficient \cite{18} for
clustering.

\noindent\textbf{Lemma 1.} \itshape If $P_1=(p'_{ij})$ and
$P_2=(p''_{ij})$ are two $n\times n$ stochastic matrices having the
inter-cluster common influence with respect to the same clustering
$\mathcal{C}$, then
$$
\tau_{\mathcal{C}}(P_1P_2)\le\tau_{\mathcal{C}}(P_1)\tau_{\mathcal{C}}(P_2)\le\min\{\tau_{\mathcal{C}}(P_1),\tau_{\mathcal{C}}(P_2)\}.
$$
\normalfont

\noindent\textbf{Proof.} We only need to show the first inequality.
Suppose that $\mathcal{C}=\{\mathcal{C}_1,\cdots,\mathcal{C}_K\}$.
We first recall a useful lemma (see \cite[p. 126, Lem 1.1]{20}): For
any stochastic matrix $P=(p_{ij})\in\mathbb{R}^{n\times n}$ and
$i,j\in V=\{1,\cdots,n\}$,
\begin{equation}
\frac12\sum_{s=1}^n|p_{is}-p_{js}|=\max_{A\subseteq V}\sum_{s\in
A}(p_{is}-p_{js}).\label{5}
\end{equation}
It follows immediately from (\ref{5}) that
$$
\tau_{\mathcal{C}}(P_1P_2)=\max_{1\le k\le
K}\max_{i,j\in\mathcal{C}_k}\max_{A\subseteq V}\sum_{s\in
A}\sum_{l=1}^n(p'_{il}-p'_{jl})p''_{ls}.
$$

Denote by $f^+=\max\{f,0\}$ and $f^-=-\min\{f,0\}$ for
$f\in\mathbb{R}$. Hence, $f=f^+-f^-$ and $|f|=f^++f^-$. Fix $1\le
k\le K$ and $i,j\in\mathcal{C}_k$. For any $1\le k'\le K$, we have
$0=\sum_{l\in\mathcal{C}_{k'}}(p'_{il}-p'_{jl})=\sum_{l\in\mathcal{C}_{k'}}(p'_{il}-p'_{jl})^+-\sum_{l\in\mathcal{C}_{k'}}(p'_{il}-p'_{jl})^-$
since $P_1$ has inter-cluster common influence. Accordingly,
$$
\sum_{l\in\mathcal{C}_{k'}}(p'_{il}-p'_{jl})^+=\sum_{l\in\mathcal{C}_{k'}}(p'_{il}-p'_{jl})^-=\frac12\sum_{l\in\mathcal{C}_{k'}}|p'_{il}-p'_{jl}|.
$$
In view of this relation, we obtain
\begin{eqnarray*}
&&\sum_{s\in A}\sum_{l=1}^n(p'_{il}-p'_{jl})p''_{ls}\\
&=&\sum_{s\in
A}\sum_{1\le k'\le
K}\sum_{l\in\mathcal{C}_{k'}}(p'_{il}-p'_{jl})p''_{ls}\\
&=&\sum_{1\le k'\le
K}\sum_{l\in\mathcal{C}_{k'}}(p'_{il}-p'_{jl})^+\sum_{s\in
A}p''_{ls}\\
&&-\sum_{1\le k'\le
K}\sum_{l\in\mathcal{C}_{k'}}(p'_{il}-p'_{jl})^-\sum_{s\in
A}p''_{ls}\\
&\le&\sum_{1\le k'\le
K}\Big(\frac12\sum_{l\in\mathcal{C}_{k'}}|p'_{il}-p'_{jl}|\Big)\max_{l\in\mathcal{C}_{k'}}\sum_{s\in
A}p''_{ls}\\
&&-\sum_{1\le k'\le
K}\Big(\frac12\sum_{l\in\mathcal{C}_{k'}}|p'_{il}-p'_{jl}|\Big)\min_{l\in\mathcal{C}_{k'}}\sum_{s\in
A}p''_{ls}\\
&\le&\sum_{1\le k'\le
K}\Big(\frac12\sum_{l\in\mathcal{C}_{k'}}|p'_{il}-p'_{jl}|\Big)\max_{l,l'\in\mathcal{C}_{k'}}\sum_{s\in
A}(p''_{ls}-p''_{l's})\\
&\le&\Big(\frac12\sum_{l=1}^n|p'_{il}-p'_{jl}|\Big)\cdot\max_{1\le
k'\le K}\max_{l,l'\in\mathcal{C}_{k'}}\sum_{s\in
A}(p''_{ls}-p''_{l's}).\\
\end{eqnarray*}

Therefore,
\begin{align}
\tau_{\mathcal{C}}(P_1P_2)\le&\Big(\max_{1\le k\le
K}\max_{i,j\in\mathcal{C}_k}\frac12\sum_{l=1}^n|p'_{il}-p'_{jl}|\Big)\nonumber\\
&\cdot\Big(\max_{A\subseteq V}\max_{1\le k'\le
K}\max_{l,l'\in\mathcal{C}_{k'}}\sum_{s\in
A}(p''_{ls}-p''_{l's})\Big),\nonumber
\end{align}
where the first term on the right-hand side is exactly
$\tau_{\mathcal{C}}(P_1)$, while the second term on the right-hand
side equals $\frac12\max_{1\le k'\le
K}\max_{l,l'\in\mathcal{C}_{k'}}\sum_{s=1}^n|p''_{ls}-p''_{l's}|=\tau_{\mathcal{C}}(P_2)$
by employing (\ref{5}). This completes the proof. $\Box$

\section{Cluster Consensus Analysis}

For a compact set $\mathcal{P}$ of $n\times n$ stochastic matrices,
we have the following assumptions:

\noindent\textbf{Assumption 1.} For all $P=(p_{ij})\in\mathcal{P}$,
$p_{ii}>0$ for $i\in V$. This means that each vertex in the graph
$G(P)$ has a self-loop.

\noindent\textbf{Assumption 2.} For each $P=(p_{ij})\in\mathcal{P}$,
if $p_{ij}>0$ then $p_{ji}>0$. Namely, $G(P)$ is an undirected
graph.

\noindent\textbf{Assumption 3.} There exists some $\delta>0$ such
that $\min\{p_{ij}:P=(p_{ij})\in\mathcal{P},p_{ij}>0\}\ge\delta.$

\noindent\textbf{Assumption 4.} For each $P=(p_{ij})\in\mathcal{P}$,
$P$ is a doubly stochastic matrix, namely,
$\sum_{i}p_{ij}=\sum_{j}p_{ij}=1$.

\noindent\textbf{Remark 3.} The positive diagonal condition in
Assumption 1 is widely adopted in the existing consensus algorithms,
see e.g. \cite{1,2,5,9}. It reflects the ``self-confidence'' that
agents give positive weights to their own states when updating
\cite{22,26}, and is also naturally satisfied by any algorithm
producing the sampling of a continuous-time process. The undirected
graph condition in Assumption 2 plays an important role in a range
of consensus problems, where information exchange goes in both
directions \cite{23}. Instances of stochastic matrices with
uniformly bounded positive entries (Assumption 3) have been
extensively investigated in discrete-time consensus with
time-varying topologies \cite{9,10,22}. If $\mathcal{P}$ is finite,
this condition is automatically satisfied. The doubly stochastic
property in Assumption 4 is important for many cooperative control
problems including distributed averaging, optimization, and
gossiping \cite{21}. A characterization of directed graphs with
doubly stochastic adjacency matrix was provided in \cite{24}.

\subsection{Equivalence Lemma}

A key step towards our main result is the following equivalence
lemma, which characterizes the cluster consensus set under the above
assumptions, and can be seen as a ``clustering'' version of Lemma
2.8 \cite{6}.

\noindent\textbf{Lemma 2.} \itshape Let $\mathcal{P}$ be a compact
set of $n\times n$ stochastic matrices having the inter-cluster
common influence with respect to the same clustering
$\mathcal{C}=\{\mathcal{C}_1,\cdots,\mathcal{C}_K\}$. Suppose that
either Assumptions 1, 2, 3 hold or Assumptions 1, 4 hold. The
following are equivalent:
\begin{itemize}
\item[(1)] $\mathcal{P}$ is a cluster consensus set.
\item[(2)] For every infinite sequence $P(1),P(2),\cdots\in\mathcal{P}$,
$\lim_{t\rightarrow\infty}P(t)P(t-1)\cdots
P(1)=\sum_{k=1}^K{\bf1}_{\mathcal{C}_k}y_k^T$, where $y_k$ is some
stochastic vector.
\item[(3)] For any $\varepsilon>0$, there is an integer
$t(\varepsilon)$ such that if $\Pi$ is the product of
$t(\varepsilon)$ matrices from $\mathcal{P}$, then
$\tau_{\mathcal{C}}(\Pi)<\varepsilon$.
\item[(4)] For all $1\le k\le K$, $i,j\in \mathcal{C}_k$, and infinite
sequences $P(1),P(2),\cdots\in\mathcal{P}$,
$\lim_{t\rightarrow\infty}(e_i^T-e_j^T)P(1)P(2)\cdots P(t)=0$.
\item[(5)] There do not exist $1\le k\le K$, $i,j\in \mathcal{C}_k$, and an infinite
sequence $P(1),P(2),\cdots\in\mathcal{P}$ such that
$e_i^TP(1)P(2)\cdots P(t)$ and $e_j^TP(1)P(2)\cdots P(t)$ have
disjoint supports for all $t\ge1$.
\item[(6)] For all infinite
sequences $P(1),P(2),\cdots\in\mathcal{P}$, $1\le k\le K$, and two
stochastic vectors $y_1$ and $y_2$ whose supports are within
$\mathcal{C}_k$,
$\lim_{t\rightarrow\infty}(y_1^T-y_2^T)P(1)P(2)\cdots P(t)=0$
\end{itemize}
\normalfont

\noindent\textbf{Proof.} The relations (1) $\Leftrightarrow$ (2),
(3) $\Rightarrow$ (4) $\Rightarrow$ (5), and (6) $\Rightarrow$ (4)
are obvious.

(2) $\Rightarrow$ (3): This can be proved by contradiction. Suppose
that (3) fails. In the light of Lemma 1, there must be some
$\varepsilon>0$ such that for every $i=1,2,\cdots$, there exists a
product $\Pi_i:=P_{i,i}P_{i,i-1}\cdots P_{i,1}$ of $i$ matrices with
$\tau_{\mathcal{C}}(\Pi_i)>\varepsilon$, where
$P_{i,j}\in\mathcal{P}$ for all $1\le j\le i$. Since $\mathcal{P}$
is compact, we choose a subsequence from $\{P_{i,1}\}_{i\ge1}$, with
the index set signified by $I_1$, so that it converges to some
accumulation point $P_1\in\mathcal{P}$. Next, we choose a new
subsequence from $\{P_{i,2}\}_{i\in I_1}$, with the index set
signified by $I_2$, so that it converges to some
$P_2\in\mathcal{P}$. By repeating this procedure, we obtain a
sequence of stochastic matrices $P_1,P_2,\cdots,$ so that, for every
integer $\ell$, we have matrices
$\Delta_1,\Delta_2,\cdots,\Delta_{\ell}$ sufficiently close to zero
satisfying $\tau_{\mathcal{C}}(Q(t')\cdots
Q(\ell+1)(P_{\ell}+\Delta_{\ell})\cdots
(P_2+\Delta_2)(P_1+\Delta_1))>\varepsilon$, where $t'\ge\ell+1$ and
$Q(t'),\cdots,Q(\ell+1)$ are some stochastic matrices. It follows
from Lemma 1 and the continuity of $\tau_{\mathcal{C}}(\cdot)$ that
$\tau_{\mathcal{C}}(P_{\ell}\cdots P_2P_1)>\varepsilon$. Clearly,
the sequence $P_1,P_2,\cdots$ does not meet the statement of (2),
which is a contradiction. This establishes the relation (2)
$\Rightarrow$ (3).

(3) $\Rightarrow$ (2): Given any infinite sequence
$P(1),P(2),\cdots\in\mathcal{P}$. If Assumptions 1, 2, and 3 hold,
the limit $\lim_{t\rightarrow\infty}P(t)P(t-1)\cdots P(1)$ exists
\cite[Thm. 2]{22}. Therefore, the statement (3) implies that the
limit must have identical rows for each cluster. Hence, (2) is true.
On the other hand, suppose that Assumptions 1 and 4 hold. Notice
that a doubly stochastic matrix is cut-balanced (see Remark 5).
Thus, Theorem 3 in \cite{26} implies that the product
$P(t)P(t-1)\cdots P(1)$ has a limit. An application of (3) again
yields (2).

(4) $\Rightarrow$ (3): Suppose that (3) does not hold. In view of
Lemma 1, there exists $\varepsilon>0$ such that for every
$i=1,2,\cdots$, there exists a product
$\Pi_i:=P_{i,i}P_{i,i-1}\cdots P_{i,1}$ of $i$ matrices with
$\tau_{\mathcal{C}}(\Pi_i)>\varepsilon$, where
$P_{i,j}\in\mathcal{P}$ for all $1\le j\le i$. Since $\mathcal{P}$
is compact, we choose a subsequence from $\{P_{i,i}\}_{i\ge1}$, with
the index set signified by $I_1$, so that it converges to some
$P_1\in\mathcal{P}$. We then choose a new subsequence from
$\{P_{i,i-1}\}_{i\in I_1}$, with the index set signified by $I_2$,
so that it converges to some $P_2\in\mathcal{P}$. By repeating this
procedure, we have a sequence of stochastic matrices
$P_1,P_2,\cdots,$ so that, for every integer $\ell$, we can choose
matrices $\Delta_1,\Delta_2,\cdots,\Delta_{\ell}$ sufficiently close
to zero satisfying
$\tau_{\mathcal{C}}((P_1+\Delta_1)\cdots(P_{\ell}+\Delta_{\ell})Q(\ell+1)\cdots
Q(t'))>\varepsilon$, where $t'\ge\ell+1$ and
$Q(t'),\cdots,Q(\ell+1)$ are some stochastic matrices. Thus,
$\tau_{\mathcal{C}}(P_1P_2\cdots P_{\ell})>\varepsilon$ by employing
Lemma 1 and the continuity of $\tau_{\mathcal{C}}(\cdot)$. By
(\ref{4}), this means that there are $1\le k\le K$, and $i(\ell),
j(\ell)\in\mathcal{C}_k$ such that
$\|(e_{i(\ell)}^T-e_{j(\ell)}^T)P_1\cdots
P_{\ell}\|_1>2\varepsilon$. Consequently, we have
$\|(e_i^T-e_j^T)P_1\cdots P_{\ell}\|_1>2\varepsilon$, where $(i,j)$
appears infinitely often in the set
$\{(i(\ell),j(\ell)):\ell=1,2,\cdots\}$. This contradicts item (4).

(5) $\Rightarrow$ (4): From item (5) we see that for any $1\le k\le
K$, $i,j\in \mathcal{C}_k$, and infinite sequences
$P(1),P(2),\cdots\in\mathcal{P}$ there exists an integer $\ell$ such
that the supports of the $i$th and $j$th rows of $P(1)P(2)\cdots
P(\ell)$ have nonempty intersection. We claim that there further
exists a uniform $\ell'$ such that for all sequences
$P(1),P(2),\cdots\in\mathcal{P}$ of length $\ell'$ the following
statement is true: For any $1\le k\le K$, $i,j\in \mathcal{C}_k$,
the supports of the $i$th and $j$th rows of $P(1)P(2)\cdots
P(\ell')$ have nonempty intersection. Indeed, if there is no such
$\ell'$, then for any $s=1,2,\cdots$ we can find a product of $s$
matrices from $\mathcal{P}$, which has two rows in some
$\mathcal{C}_{k'}$ whose supports do not intersect. By similar
argument used in the proof of ``(4) $\Rightarrow$ (3)'', we have
$\tau_{\mathcal{C}}(P_1P_2\cdots P_s)=1$ involving Lemma 1 and the
continuity. Hence, there must exist $1\le k''\le K$ and
$i,j\in\mathcal{C}_{k''}$ such that the $i$th and $j$th rows of
$P_1P_2\cdots P_s$ have disjoint supports for infinitely many $s$.
Notice that the initial statement says that there is an integer
$\ell$ satisfying $\tau_{\mathcal{C}}(P_1\cdots P_{\ell})<1$. But
now we can pick $\bar{\ell}\ge\ell+1$ such that
$\tau_{\mathcal{C}}(P_1\cdots P_{\ell}\cdots P_{\bar{\ell}})=1$. We
reach a situation that is at odds with Lemma 1. This proves the
claim.

Fix the above obtained $\ell'$. For an integer s, denote by
$\Pi_s:=P(1)\cdots P(s)$. Define
$\beta:=\sup\{\tau_{\mathcal{C}}(\Pi_{\ell'}):P(1),\cdots,$
$P(\ell')$ $\in\mathcal{P}\}$. It is clear that $0\le\beta\le1$. If
$\beta=1$, then there must exist some product $P(1)\cdots P(\ell')$
which has two rows within the same cluster having disjoint supports
since any continuous function on a compact set attains its supremum.
This contradicts the above claim. Hence,
$\tau_{\mathcal{C}}(\Pi_{\ell'})\le\beta<1$ for any $\Pi_{\ell'}$
from $\mathcal{P}$. For any $\varepsilon>0$, by taking
$m=\lceil\ln_{\beta}\varepsilon\rceil+1$ and $t=m\ell'$ we have
$\tau_{\mathcal{C}}(\Pi_t)=\tau_{\mathcal{C}}(\Pi_{m\ell'})\le\beta^m<\varepsilon$.
Thus, item (3) holds and (4) follows.

(4) $\Rightarrow$ (6): Given any $1\le k\le K$, and stochastic
vectors $y_1$ and $y_2$ whose supports are within $\mathcal{C}_k$.
Without loss of generality, we assume $k=1$ and
$\mathcal{C}_1=\{1,2,\cdots,n_1\}$ $(1\le n_1\le n)$. Since
$\{e_i-e_{i+1}: i=1,\cdots,n_1-1\}$ is a base of the subspace of
$\mathbb{R}^{n_1}$ that is orthogonal to ${\bf1}_{\mathcal{C}_1}$,
we have $y_1-y_2=\sum_{i=1}^{n_1-1}a_i(e_i-e_{i+1})=\sum_{i<j,\ i,
j\in\mathcal{C}_1}a_{ij}(e_i-e_j)$ for some numbers $a_i$ and
$a_{ij}$. Therefore, for any infinite sequence
$P(1),P(2),\cdots\in\mathcal{P}$,
$\lim_{t\rightarrow\infty}(y_1^T-y_2^T)P(1)\cdots P(t)=\sum_{i<j,\
i,
j\in\mathcal{C}_1}a_{ij}\lim_{t\rightarrow\infty}(e_i^T-e_j^T)P(1)\cdots
P(t)=0$. $\Box$

\noindent\textbf{Remark 4.} In Lemma 2, Assumptions 1, 2, 3, and 4
are only used in ``(3) $\Rightarrow$ (2)''. The equivalence of (1)
and (5) will be critical in our following combinatorial
characterization of cluster consensus set. As such, even without the
four assumptions, we still have ``(1) $\Rightarrow$ (5)''.

\noindent\textbf{Remark 5.} We mention that the doubly stochasticity
in Assumption 4 can be replaced with a weaker (but more
sophisticated) condition, which is called \textit{cut-balance}
\cite{23,25}. $P=(p_{ij})$ is cut-balanced if there exists $C\ge1$
such that for every $S\subseteq V$, $\sum_{i\in S}\sum_{j\in
V\backslash S}p_{ij}\le C\sum_{i\in V\backslash S}\sum_{j\in
S}p_{ij}$.

\subsection{Avoiding Sets Condition}

Given $P,P(1),P(2),\cdots\in\mathcal{P}$ and $S\subseteq V$, we will
write $N_P(S):=N_{G(P)}(S)$, $N_i(S):=N_{G(P(i))}(S)$,
$N^1(S):=N_1(S)$, $N^2(S):=N_2(N_1(S))$, etc. following \cite{6} for
ease of notation. The ``clustering'' version of the avoiding sets
condition is as follows.

\noindent\textbf{Theorem 1.} \itshape Let $\mathcal{P}$ be a compact
set of $n\times n$ stochastic matrices having the inter-cluster
common influence with respect to the same clustering
$\mathcal{C}=\{\mathcal{C}_1,\cdots,\mathcal{C}_K\}$. Suppose that
either Assumptions 1, 2, 3 hold or Assumptions 1, 4 hold.
$\mathcal{P}$ is not a cluster consensus set if and only if there
exist two sequences of nonempty subsets of $V$,
$$
S_1, S_2, \cdots, S_\ell\quad and\quad S'_1, S'_2, \cdots, S'_\ell,
$$
of length $\ell\le3^n-2^{n+1}+1$ and a sequence of matrices
$P(1),P(2),\cdots,P(\ell)\in\mathcal{P}$ satisfying (i) $S_l\cap
S'_l=\emptyset$, $l=1,\cdots,\ell$; (ii) For any integer $s\ge0$,
$N_l(S_l)\subseteq S_{l+1}$, $l\equiv s(\mod\ell)+1$; and (iii)
There exist $i\in S_1$ and $j\in S'_1$ such that
$i,j\in\mathcal{C}_k$ for some $1\le k\le K$. \normalfont

\noindent\textbf{Remark 6.} The condition (iii) is an essential
difference as compared to Theorem 2.2 \cite{6}. Since the two
sequences are set cycles, $S_1$ and $S'_1$ in (iii) can well be
substituted by any pair of $S_l$ and $S'_l$. Roughly speaking,
Theorem 1 says that $\mathcal{P}$ is a cluster consensus set if and
only if there do not exist two set cycles which are disjoint at
every step and contain vertices from the same cluster at some step.

\noindent\textbf{Remark 7.} It is clear that, under the assumptions
of Theorem 1, deciding whether a finite set $\mathcal{P}$ of
stochastic matrices is a cluster consensus set is algorithmically
decidable (c.f. \cite[Prop. 2.6]{6}).

To see how Theorem 1 can be used, some examples are in order before
we plunge into the proof.

\noindent\textbf{Example 1.} Assume that $P=(p_{ij})$ is an $n\times
n$ stochastic matrix having positive diagonal and the inter-cluster
common influence with respect to a clustering
$\mathcal{C}=\{\mathcal{C}_1,\cdots,\mathcal{C}_K\}$. Suppose
further that either $G(P)$ is undirected or $P$ is doubly
stochastic. If there is some $1\le k\le K$ such that $\mathcal{C}_k$
can be partitioned into two disjoint nonempty sets $U_1$ and $U_2$
with $p_{ij}=p_{ji}=0$ for all $i\in U_1, j\in U_2$ and $\sum_{j\in
U_1}p_{i_1,j}=\sum_{j\in U_2}p_{i_2,j}=1$ for some $i_1\in U_1$ and
$i_2\in U_2$, then the singleton $\mathcal{P}=\{P\}$ obviously is
not a cluster consensus set. Indeed, we can take $S_1=U_1$,
$S'_1=U_2$, and $\ell=1$. The inter-cluster common influence
condition implies that $N_P(S_1)\subseteq S_1$ and
$N_P(S'_1)\subseteq S'_1$.

\noindent\textbf{Example 2.} Suppose $\mathcal{P}$ is a compact set
of $n\times n$ stochastic matrices having the inter-cluster common
influence with respect to the same clustering
$\mathcal{C}=\{\mathcal{C}_1,\cdots,\mathcal{C}_K\}$. In addition,
either Assumptions 1, 2, 3 hold or Assumptions 1, 4 hold. If for all
$P\in\mathcal{P}$ the induced subgraphs of $G(P)$ on $\mathcal{C}_k$
for all $1\le k\le K$ are strongly connected, then $\mathcal{P}$ is
a cluster consensus set. This can be justified by Theorem 1 as
follows. For any sequence of matrices
$P(1),P(2),\cdots\in\mathcal{P}$ and any pair of subsets $S_1$ and
$S'_1$ such that there are $i\in S_1$, $j\in S'_1$, and
$i,j\in\mathcal{C}_k$ for some $1\le k\le K$, we must have
$\mathcal{C}_k\subseteq N^{n-1}(S_1)$ and $\mathcal{C}_k\subseteq
N^{n-1}(S'_1)$. Thus, two avoiding set cycles cannot exist.

\noindent\textbf{Example 3.} Suppose that $P$ is an $n\times n$
stochastic matrix having positive diagonal and the inter-cluster
common influence with respect to a clustering
$\mathcal{C}=\{\mathcal{C}_1,\cdots,\mathcal{C}_K\}$. Suppose
further that either $G(P)$ is undirected or $P$ is doubly
stochastic. If $G(P)$ has \textit{cluster-spanning trees} with
respect to $\mathcal{C}$ (i.e., for each cluster $\mathcal{C}_k$,
$1\le k\le K$, there is a vertex $i_k\in V$ such that there exist
paths in $G(P)$ from all vertices in $\mathcal{C}_k$ to $i_k$), then
$\{P\}$ is a cluster consensus set \cite[Thm 1]{9}. This can be
deduced quickly from Theorem 1. Indeed, for any pair of subsets
$S_1$ and $S'_1$ such that there are $i\in S_1$, $j\in S'_1$, and
$i,j\in\mathcal{C}_{k'}$ for some $1\le k'\le K$, we obtain $i_k\in
N^{n-1}(S_1)\cap N^{n-1}(S'_1)$. Clearly, two avoiding set cycles
cannot occur in this case.

\noindent\textbf{Example 4.} Example 3 can be generalized to tackle
switching topologies. Consider the infinite products of stochastic
matrices $\cdots Q(2)Q(1)$ such that $\{Q(t):t\ge1\}$ have the
inter-cluster common influence with respect to the same clustering
$\mathcal{C}$. Assume that (i) either Assumptions 1, 2, 3 hold or
Assumptions 1, 4 hold for $\{Q(t):t\ge1\}$; and (ii) the graph
obtained by joining the edge sets of the graphs
$G(Q(iL+1)),\cdots,G(Q((i+1)L))$ contains cluster-spanning trees
with respect to $\mathcal{C}$ for every integer $i\ge0$. Then, the
dynamic system $x(t+1)=Q(t+1)x(t)$ achieves cluster consensus (c.f.
\cite[Thm. 3]{9} and \cite[Thm. 2]{10}). We can derive this from
Theorem 1 by first noting that the product $P(i):=Q((i+1)L)\cdots
Q(iL+1)$ has the inter-cluster common influence with respect to
$\mathcal{C}$, and it still satisfies Assumptions 1, 2, 3 or
Assumptions 1, 4 (according to whether the former or the latter
holds in (i)). Define $\mathcal{P}:=\{P(i):i\ge0\}$. The same
reasoning in Example 3 implies that two avoiding set cycles cannot
occur. Hence, the system $x(t+1)=Q(t+1)x(t)$ reaches cluster
consensus.

\noindent\textbf{Example 5.} An interesting implication of Theorem 1
is the following result, which says that having cluster-spanning
trees is also necessary for cluster consensus.

\noindent\textbf{Corollary 1.} \itshape Let $\mathcal{P}$ be a
compact set of $n\times n$ stochastic matrices having the
inter-cluster common influence with respect to the same clustering
$\mathcal{C}=\{\mathcal{C}_1,\cdots,\mathcal{C}_K\}$. Suppose that
either Assumptions 1, 2, 3 hold or Assumptions 1, 4 hold. If
$\mathcal{P}$ is a cluster consensus set, then, for any
$P\in\mathcal{P}$, $G(P)$ has cluster-spanning trees with respect to
$\mathcal{C}$. \normalfont

\noindent\textbf{Proof.} If $\mathcal{P}$ is a cluster consensus
set, then so it is with $\{P\}$ for any $P\in\mathcal{P}$. The
condensation of $G(P)$, denoted by $CG(P)$, is a directed acyclic
graph formed by contracting the strongly connected components of
$G(P)$. It is well known that the condensation of a graph has at
least one sink, i.e., a vertex with no out-neighbors. We claim that
if $CG(P)$ has at least two sinks, then we cannot find two sinks
both containing vertices in the same cluster.

Indeed, if this is not true, then we obtain two nodes $Sink_1$ and
$Sink_2$ from $CG(P)$ such that $i\in Sink_1$ and $j\in Sink_2$ for
some $i,j\in\mathcal{C}_k$ and $1\le k\le K$. Take $S_1=Sink_1$ and
$S'_1=Sink_2$. Then $S_1\cap S'_1=\emptyset$, $N_P(S_1)\subseteq
S_1$, and $N_P(S'_1)\subseteq S'_1$. It follows from Theorem 1 that
$\{P\}$ is not a cluster consensus set, which contradicts our
assumption. This establishes the claim.

Therefore, if $CG(P)$ has precisely one sink, then each vertex in
this sink represents the root of a spanning tree of $G(P)$. Of
course, $G(P)$ has cluster-spanning trees with respect to
$\mathcal{C}$. If $CG(P)$ has at least two sinks, by our above
claim, for any cluster $\mathcal{C}_k$, the vertices of
$\mathcal{C}_k$ lie in at most one sink (some vertices of
$\mathcal{C}_k$ may lie in non-sink nodes of $CG(P)$). It is easy to
see that $G(P)$ has cluster-spanning trees with respect to
$\mathcal{C}$ in this case. $\Box$

\noindent\textbf{Proof of Theorem 1.} \textit{Sufficiency.} Suppose
the sequences of sets $S_1,\cdots,S_{\ell}$,
$S'_1,\cdots,S'_{\ell}$, and
$P(1),P(2),\cdots,P(\ell)\in\mathcal{P}$ satisfying (i), (ii), and
(iii) exist. Consider the infinite sequence of matrices made up of
$P(1),P(2),\cdots,P(\ell)$ occurring periodically in this order. In
the light of (iii), we pick $i\in S_1$ and $j\in S'_1$ such that
$i,j\in\mathcal{C}_k$ for some $1\le k\le K$. Now we claim that the
two vectors $e_i^TP(1)\cdots P(t)$ and $e_j^TP(1)\cdots P(t)$ have
disjoint supports for all $t\ge1$. Therefore, $\mathcal{P}$ is a
cluster consensus set by Lemma 2 ``(1) $\Rightarrow$ (5)''.

It remains to show the claim. Indeed, for any $t\ge1$, the support
of $e_i^TP(1)\cdots P(t)$ is just $N^t(\{i\})$, which is contained
in $S_{t(\mod \ell)+1}$. Likewise, the support of $e_j^TP(1)\cdots
P(t)$ is $N^t(\{j\})$, which is contained in $S'_{t(\mod \ell)+1}$.
But by assumption, $S_{t(\mod \ell)+1}\cap S'_{t(\mod
\ell)+1}=\emptyset$. The proves the claim.

\textit{Necessity.} We will show the necessity by contradiction.
Suppose that there are no such sets $S_1,\cdots,S_{\ell}$,
$S'_1,\cdots,S'_{\ell}$, and
$P(1),P(2),\cdots,P(\ell)\in\mathcal{P}$ exist. Take any sequence
$Q(1),Q(2),\cdots,\in\mathcal{P}$. Pick $i,j\in\mathcal{C}_k$ for
some $1\le k\le K$, and define $U_t:=N^t(\{i\})$ and
$U'_t:=N^t(\{j\})$ for $t\ge1$. We claim that there must exist some
$t_0\ge1$ such that $U_{t_0}\cap U'_{t_0}\not=\emptyset$. Since
$U_t$ is the support of $e_i^TQ(1)\cdots Q(t)$ and $U'_t$ is the
support of $e_j^TQ(1)\cdots Q(t)$, the claim would imply that
$\mathcal{P}$ is a cluster consensus set by using Lemma 2 ``(5)
$\Rightarrow$ (1)''. This contradicts our assumption and completes
the proof of necessity.

To show the claim, we again assume the opposite. Suppose that, for
any $t\ge1$, $U_t\cap U'_t=\emptyset$. There exist two integers
$a<b$ such that $(U_a,U'_a)=(U_b,U'_b)$. Define $S_1:=U_a$,
$S'_1:=U'_a$, $P(1):=Q(a)$, $P(2):=Q(a+1)$, $\cdots,
P(\ell):=Q(\ell+a-1)$, where $\ell:=b-a$. Let $S_{l+1}:=N_l(S_l)$
and $S'_{l+1}:=N_l(S'_l)$ for $1\le l\le \ell$. Thereby we get two
avoiding set cycles satisfying (i) and (ii) of Theorem 1. Since
$P(1),\cdots,P(\ell)$ all have positive diagonals, we see that $i\in
S_1$, $j\in S'_1$ and hence item (iii) is also satisfied. Finally,
by a basic combinatorial outcome that the number of ordered
partitions $(A,B,C)$ of $V$ with nonempty $A, B$ and empty
intersection of any two of them is $3^n-2^{n+1}+1$ (see e.g.
\cite[p. 90]{27}), we have $\ell\le3^n-2^{n+1}+1$ by deleting some
possible repetitions. This is at odds with our initial assumption,
which in turn proves the claim. $\Box$

We observe that, without Assumptions 1, 2, 3, 4, the proof for
sufficiency of Theorem 1 above still holds by recalling Remark 4.
Therefore, we obtain the following necessary condition for cluster
consensus without requiring any of the four assumptions.

\noindent\textbf{Corollary 2.} \itshape Let $\mathcal{P}$ be a
compact set of $n\times n$ stochastic matrices having the
inter-cluster common influence with respect to the same clustering
$\mathcal{C}=\{\mathcal{C}_1,\cdots,\mathcal{C}_K\}$. If
$\mathcal{P}$ is a cluster consensus set, then there do not exist
two sequences of nonempty subsets of $V$,
$$
S_1, S_2, \cdots, S_\ell\quad and\quad S'_1, S'_2, \cdots, S'_\ell,
$$
of length $\ell\le3^n-2^{n+1}+1$ and a sequence of matrices
$P(1),P(2),\cdots,P(\ell)\in\mathcal{P}$ satisfying (i) $S_l\cap
S'_l=\emptyset$, $l=1,\cdots,\ell$; (ii) For any integer $s\ge0$,
$N_l(S_l)\subseteq S_{l+1}$, $l\equiv s(\mod\ell)+1$; and (iii)
There exist $i\in S_1$ and $j\in S'_1$ such that
$i,j\in\mathcal{C}_k$ for some $1\le k\le K$. \normalfont

\section{Conclusions}

In this technical note, we have presented a combinatorial necessary
and sufficient condition for cluster consensus of discrete-time
linear systems. This combinatorial condition can be thought of as an
extension of the original avoiding sets condition (i.e., $K=1$) that
was shown to be responsible for global consensus \cite{6}. The
result can be used to show that the cluster-spanning trees condition
proposed in \cite{9} is not only sufficient but necessary in some
sense for achieving cluster consensus.

Note that the concept of cluster consensus in our paper is built on
an underlying fixed clustering of the networks. In a general
context, the multi-agent system may have changing clusterings over
time. How to extend the presented method to deal with dynamical
clustering is very appealing. Another direction worth investigation
is the complexity of decidability of cluster consensus. In
particular, developing efficient algorithms for cluster consensus
decision problems would be imperative.

\ifCLASSOPTIONcaptionsoff
  \newpage
\fi

\end{document}